\documentclass[aps,prl,longbibliography,reprint,showpacs,floatfix,superscriptaddress]{revtex4-1}
\usepackage{graphicx}
\usepackage{amsmath,amssymb,amsfonts}
\usepackage[english]{babel}
\usepackage[bookmarks=false]{hyperref}
\hypersetup{
		breaklinks=true,
    unicode=false,          % non-Latin characters in Acrobat’s bookmarks
    pdftoolbar=true,        % show Acrobat’s toolbar?
    pdfmenubar=true,        % show Acrobat’s menu?
    pdffitwindow=false,     % window fit to page when opened
    pdfstartview={FitH},    % fits the width of the page to the window
    pdftitle={Taufour},    % title
    pdfauthor={Taufour},     % author
    pdfsubject={CeCoMg},   % subject of the document
    pdfcreator={Taufour},   % creator of the document
    pdfproducer={Taufour}, % producer of the document
    pdfkeywords={keyword1} {key2} {key3}, % list of keywords
    pdfnewwindow=true,      % links in new window
    colorlinks=true,       	% false: boxed links; true: colored links
    linkcolor=black,          % color of internal links         black for printing
    citecolor=black,    	    % color of links to bibliography  black for printing
    filecolor=black,   		    % color of file links             black for printing
    urlcolor=blue            % color of external links         black for printing
}

\begin{document}

\title{Ce$_{3-x}$Mg$_x$Co$_{9}$: transformation of a Pauli paramagnet into a strong permanent magnet}
\author{Tej N. Lamichhane} 
\affiliation{Ames Laboratory, U.S. DOE, Ames, Iowa 50011, USA}
\affiliation{Department of Physics and Astronomy, Iowa State University, Ames, Iowa 50011, USA}
\author{Valentin Taufour}
\affiliation{Ames Laboratory, U.S. DOE, Ames, Iowa 50011, USA}
\affiliation{Department of Physics, University of California Davis, Davis, California 95616, USA}
\author{Andriy Palasyuk}
\affiliation{Ames Laboratory, U.S. DOE, Ames, Iowa 50011, USA}
\author{Qisheng Lin}
\affiliation{Ames Laboratory, U.S. DOE, Ames, Iowa 50011, USA}
\author{Sergey L. Bud'ko}
\affiliation{Ames Laboratory, U.S. DOE, Ames, Iowa 50011, USA}
\affiliation{Department of Physics and Astronomy, Iowa State University, Ames, Iowa 50011, USA}
\author{Paul C. Canfield}
\affiliation{Ames Laboratory, U.S. DOE, Ames, Iowa 50011, USA}
\affiliation{Department of Physics and Astronomy, Iowa State University, Ames, Iowa 50011, USA}
%\altaffiliation{Materials Science and Technology Division, Oak Ridge National Laboratory, Oak Ridge, Tennessee 37831}
%\address[label2]{Department of Physics,University of California,Davis,CA 95616, U.S.A.}
\begin{abstract}
We report on the synthesis of single crystalline and polycrystalline samples of Ce$_{3-x}$Mg$_x$Co$_9$ solid solution ($0\leq x \lesssim 1.4$) and characterization of their structural and magnetic properties. The crystal structure remains rhombohedral in the whole composition range and Mg  partially replaces Ce in the 6\textit{c} site of the CeCo$_3$ structure. Ferromagnetism is induced by Mg substitutions starting as low as  $x=0.18$ and reaching a Curie temperature as high as $450$\,K for $x=1.35$. Measurements on single crystals with $x=1.34$ and $T_\textrm{C}=440$\,K indicate an axial magnetic anisotropy with the anisotropy field of $6$\,T and a magnetization of $6$\,$\mu_B/$f.u. at $300$\,K. Coercicity is observed in the polycrystalline samples consistent with the observed axial magnetic anisotropy. Our discovery of ferromagnetism with large axial magnetic anisotropy induced by substituting a rare-earth element by Mg is a very promising result in the search of inexpensive permanent-magnet materials and suggests other non-magnetic phases, similar to CeCo$_3$, may also conceal nearby ferromagnetic phases.
\end{abstract}

\maketitle

\section{Introduction}

Current rare-earth-based commercial magnets contain local moment bearing rare-earth elements, mainly Nd, Sm and Dy whose availability is, according to the United States  Department of Energy, important to the clean energy economy and also have an associated supply risk. Alternative to finding a long sought rare earth free, high flux permanent magnet, attempts to find Ce-based permanent magnets or substituting Ce for more critical rare-earth elements could be a pragmatic strategy to address the criticality problem since Ce is a relatively more abundant rare earth \cite{Relativeabundance} with easier extraction chemistry. Ce is relatively easy to separate from the other rare earths since it can easily be oxidised to CeO$_2$ via roasting from which it can be precipitated out in acidic solutions  \cite{PPatnaikHandbook,XIE201410}. Ce can be, in theory as well as experiment, a  substitute for critical rare-earths without much compromise in magnetic properties \cite{PathakA}. Because of the volatile price of critical rare-earths, Ce is drawing attention for developing gap-magnets which will populate the gap in energy product (in the range of 10-20 MGOe) in between low flux (Alnico, ferrites) and  commercial rare-earth based magnets such as SmCo$_{5}$ and Nd$_{2}$Fe$_{14}$B. Additionally, the study of new Ce based compounds often can reveal interesting physics; Ce exhibits diverse electronic and magnetic  properties like local moment bearing Ce$^{3+}$ ion, nonmagnetic Ce$^{4+}$ ion, mixed valency, intermediate valence and itinerant magnetism. 

 Ce$_{3-x}$Mg$_x$Co$_9$ alloys are substitution derivatives of CeCo$_3$ in which Mg partially replace Ce in the 6\textit{c} position. The hydrogenation properties of the compound Ce$_2$MgCo$_9$ ($x=1$) and Nd$_{3-x}$Mg$_x$Co$_9$ alloys ($x\leq 1.5$) have been recently investigated~\cite{Denys2012JSSC,Shtender2017JAC}. In addition, it was shown that the substitution of Nd by Mg increases the Curie temperature from $381$\,K for NdCo$_3$~\cite{Bartashevich1993JAC} to $633$\,K for Nd$_2$MgCo$_9$~\cite{Shtender2017JAC}. 

In this work we present  structural and  magnetic properties of both single crystalline and polycrystalline Ce$_{3-x}$Mg$_{x}$Co$_{9}$ for~$0\leq x\lesssim 1.4$. The anisotropic magnetic properties are studied for a single crystal of Ce$_{1.662(4)}$Mg$_{1.338(4)}$Co$_{9}$ (Hereafter we round off the single crystalline composition to 3-significant digits as Ce$_{1.66}$Mg$_{1.34}$Co$_9$.). We find a remarkable transformation of a Pauli-paramagnet CeCo$_{3}$ (Ce$_{3}$Co$_{9}$) into the potential permanent magnet Ce$_{3-x}$Mg$_x$Co$_9$ which develops $2.2$~MJ/m$^3$ of uniaxial anisotropy energy at $2$~K for the Ce$_{1.66}$Mg$_{1.34}$Co$_9$.    

\begin{figure}[!h]
\includegraphics[scale =0.35]{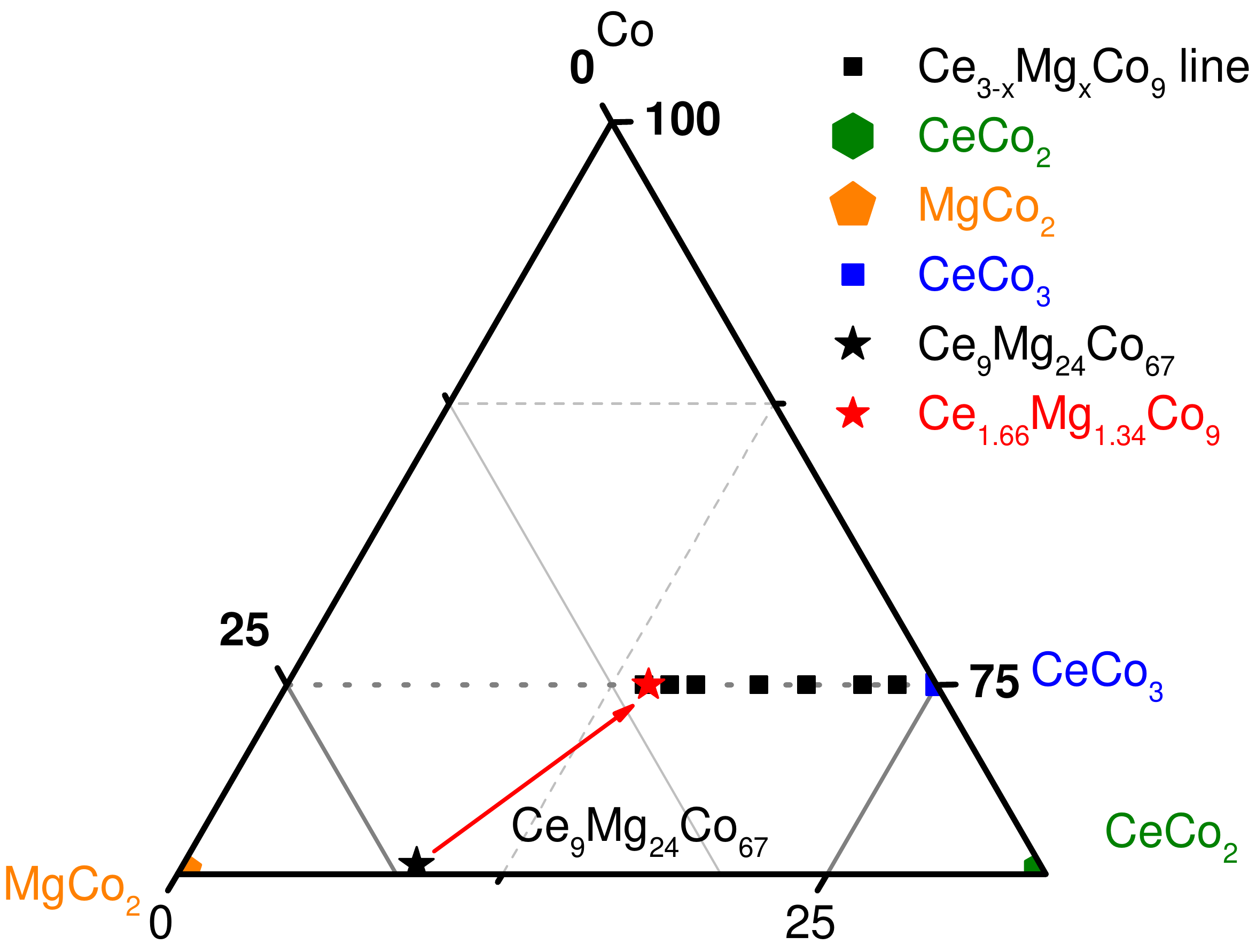}
\caption{Co-rich portion of the Co-Mg-Ce ternary phase diagram showing the Ce$_{3-x}$Mg$_{x}$Co$_{9}$ solid solution line and 1:2 type impurity phases. Ce$_9$Mg$_{24}$Co$_{9}$ is the optimised initial melt composition for the solution growth of single crystalline Ce$_{1.66}$Mg$_{1.34}$Co$_{9}$. (See text for details.)}
\label{Composition}
\end{figure}

\section{Experimental Methods}

To establish the existence range of the solid solution Ce$_{3-x}$Mg$_x$Co$_9$, we prepared polycrystals with various nominal compositions of $0\leq x\leq 2.00$ (see TABLE~\ref{tbl:EDS} below). Ce metal from the Ames Laboratory Material Preparation Center (purity $>$ 99.99\%), Co chunks ($99.95\%$, Alfa Aesar), and Mg ($99.95\%$,Gallium Source, LLC) were packed in a 3-capped Ta crucible~\cite{Canfield2001JCG} under an Ar atmosphere. The Ta crucible was then sealed into an amorphous silica ampoule. The ampoule was heated to $900~{^\circ}\mathrm{C}$ over $3$ hours and held there for $3$ hours. This step allows the reaction of Ce and Mg at low temperature and avoids the excessive boiling of Mg inside the Ta crucible. The ampoule was then heated to $1200~{^\circ}\mathrm{C}$ over $3$ hours and held there for 10 hours. At this point, the ampoule was spun in a centrifuge and all the molten growth material was decanted and quenched in catch side of the Ta-tube. This step confirmed that the mixture was forming a homogeneous melt at $1200~{^\circ}\mathrm{C}$. The ampoule was put back into a furnace then annealed at $900~{^\circ}\mathrm{C}$ for $24$ hours. Mg free CeCo$_3$ was also synthesized by arcmelting the stoichiometric composition and annealing at $900~{^\circ}\mathrm{C}$ for 1 week.

Single crystals of Ce$_{3-x}$Mg$_x$Co$_9$ were  grown using a solution growth technique. An initial composition of Ce$_9$Mg$_{24}$Co$_{67}$ (see FIG.~\ref{Composition}) was packed in a 3-capped Ta crucible~\cite{Canfield2001JCG} and heated to $1200~{^\circ}\mathrm{C}$ similar to the polycrystals. The ampoule was then cooled down to $1100~{^\circ}\mathrm{C}$ over $75$~hours after which crystals were separated from the flux by using a centrifuge. Similarly, CeCo$_{3}$ single crystals were prepared by cooling a Ce$_{30}$Co$_{70}$ melt from $1200^\circ$ to $1100~{^\circ}\mathrm{C}$ in 1 hour and then to $1050~{^\circ}\mathrm{C}$ over 75 hours\cite{ASMFCeCo}.

Elemental analysis of the samples was performed using Energy Dispersive Spectroscopy(EDS). Polycrystalline samples were embedded in epoxy resin and finely polished. The polished samples were examined with EDS on 6-10 spots and a statistical average composition is reported. Thin plate-like single crystalline samples, see FIG.~\ref{CombinedXRD}(a), were mounted on a conducting carbon tape. Self-flux grown MgCo$_2$ and CeCo$_2$  single crystals were used as absorption standards for the Ce-Co-Mg alloy composition analyses. Powder X-ray diffraction data were collected at room temperature on a Rigaku MiniFlex II diffractometer with Cu K$\alpha$ radiation. Data were collected with a $3$~seconds dwell time for each interval of $0.01^\circ$ within a 2$\theta$ range of $10-100^\circ$. Lattice parameters were refined by the Rietveld analysis method using GSAS~\cite{Larson2004} and EXPGUI~\cite{Toby2001JAC}.
% A profile $R$-factor of $R_P<0.15$ was obtained in all cases.

Single crystal X-ray diffraction was carried out on a Bruker Smart APEX II diffractometer with graphite-monochromatized Mo-\textit{K}$_\alpha$ radiation ($0.71073$~\AA). Reflections were gathered at room temperature by taking four sets of $360$~frames with $0.5^\circ$ scans in $\omega$, with an exposure time of $10$~s. The crystal-to-detector distance was maintained $60$~mm. The reflections were collected over the range of $3^\circ$ to $62^\circ$ in $2\theta$.

Electrical resistivity was measured on single crystals using the 4-probe technique with a Linear Research, Inc. ac resistance bridge (LR700, \textit{f} = 17 Hz). A Quantum Design (QD) Magnetic Property Measurement System (MPMS) was used for the temperature control. Samples were sliced into thin rectangular bars (approximately 0.9 mm x 0.45 mm x 0.04 mm) and platinum wires were attached to the samples with Dupont 4929N silver paint. The contact resistances were less than 2 $\Omega$.

Magnetization was measured using a QD-VersaLab Vibrating Sample Magnetometer (VSM). The standard option was used in the temperature range $50$-$400$~K and the oven option in the range $300$-$1000$~K. Loctite 435 and Zircar cement were used to attach the samples in the standard and oven options, respectively. Field dependent magnetization isotherms were also measured down to $2$~K in a MPMS. The details of sample mounting and experimental determination of demagnetization factor along the easy axis are discussed in references~\onlinecite{Lamichhane2015} and ~\onlinecite{HfZrMnPTej}.

\section{Composition and structural properties}
\subsubsection{Single crystal: Characterization and structure}

\begin{figure*}[!h]
\includegraphics[scale =0.75]{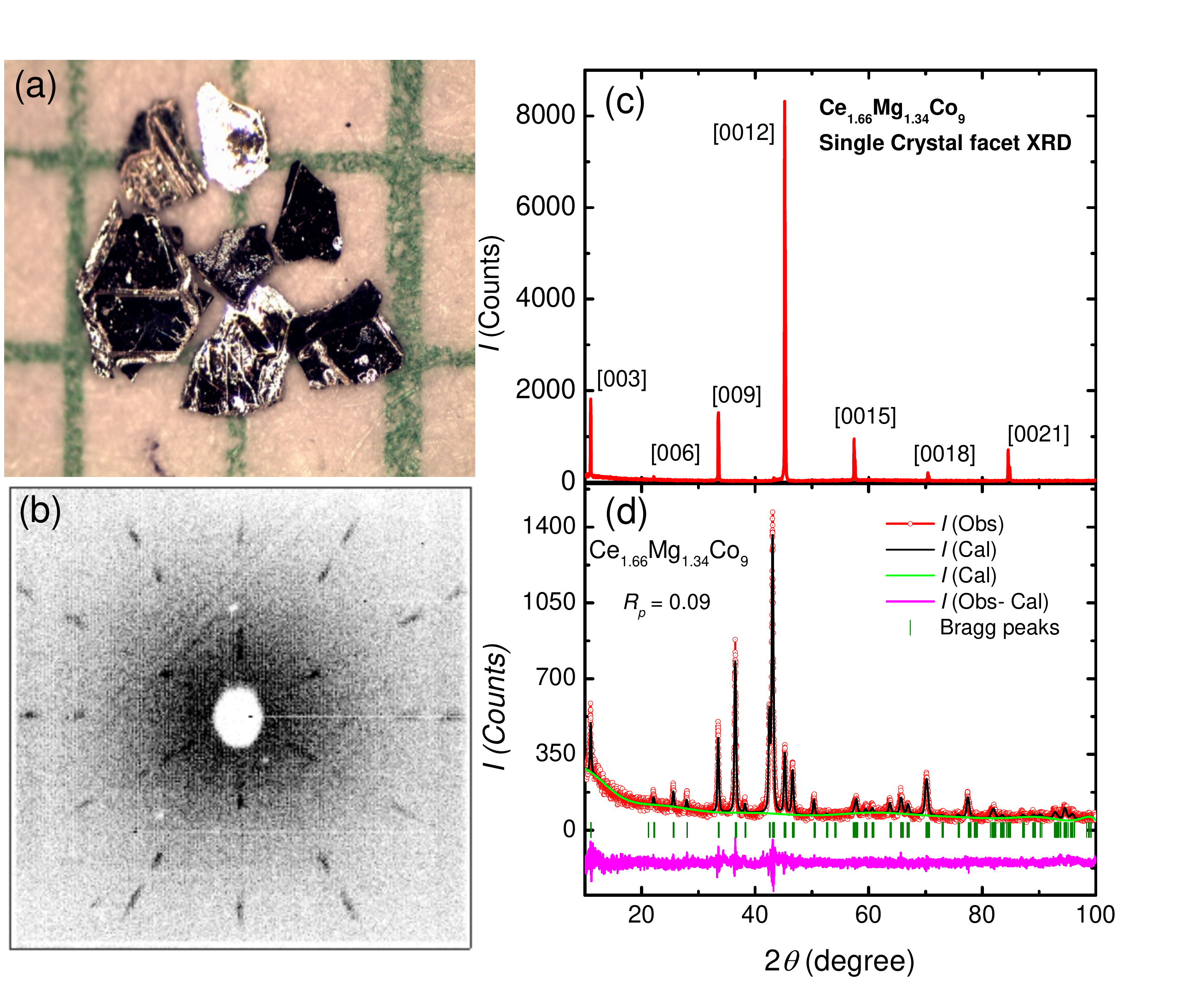}
\caption{(a) Single crystals of Ce$_{1.66}$Mg$_{1.34}$Co$_{9}$. (b) Backscattered Laue photograph of Ce$_{1.66}$Mg$_{1.34}$Co$_{9}$ with X-ray beam perpendicular to plate. (c) Monochromatic X-ray diffraction from the surface of single crystal using Bragg-Brentano geometry. (d) Powder XRD for Ce$_{1.66}$Mg$_{1.34}$Co$_{9}$ where \textit{I}(Obs), \textit{I}(Cal) and \textit{I}(Bkg) are experimental, Rietveld refined and instrumental background data respectively. The lower section of the graph shows the Bragg's peaks positions with olive vertical lines and  the differential X-ray diffractrogram \textit{I}(Obs-Cal). 
\label{CombinedXRD}}
\end{figure*}

\begin{table}[!h]
\begin{center}
\caption{Crystallographic data and refinement parameters for Ce$_{1.662(4)}$Mg$_{1.338(4)}$Co$_{9}$.}
\label{tbl:crystaldata}
\begin{tabular}{|l|l|}
\hline
Empirical formula & Ce$_{1.662(4)}$Mg$_{1.338(4)}$Co$_{9}$ \\
Formula weight & 796.32 \\
Crystal system, space group & trigonal, \textit{R-3m} h \\
Unit cell dimensions & $a = 4.9260(7)$~\AA \\
  & $c = 24.019(5)$~\AA \\
Volume & $504.75(18)$~\AA$^3$ \\
Z, Calculated density & $3$, $7.859$~g/cm$^3$ \\
Absorption coefficient & $32.577$~mm$^{-1}$ \\

Reflections collected & 2000 [R$_{(\textit{int})}$ = 0.0408] \\

Data / restraints / parameters & 224 / 0 / 18\\
Goodness-of-fit on $|F|^2$ & 1.149 \\
Final $R$ indices [$I>4\sigma(I)$] & $R1 = 0.0204$, $wR2 = 0.0450$ \\
$R$ indices (all data) & $R1 = 0.0226$, $wR2 = 0.0455$\\
Largest diff. peak and hole & 1.917 and -1.477 e.\AA$^{-3}$\\ \hline
\end{tabular}
\end{center}
\end{table}

\begin{table}[!h]
\begin{center}
\caption{Atomic coordinates and equivalent isotropic displacement parameters ($\AA^2\times10^{-3}$) for  Ce$_{1.66}$Mg$_{1.34}$Co$_{9}$. \textit{U}$_{(\textit{eq})}$ is defined as one third of the trace of the orthogonalized \textit{U}$_{\textit{ij}}$ tensor.}
\label{tbl:atomcoord}
\begin{tabular}{| p{1cm}|p{1.1cm} | p{0.75cm} | p{1.3cm} | p{1.3cm} | p{1.3cm} |p{0.85cm}|}
\hline
atom & Wyckoff site & Occ & \textit{x} & \textit{y} & \textit{z} & \textit{U}$_{\textit{eq}}$ \\ \hline
Ce(1)&3\textit{a} & 1&0	& 0 &	0 &	14(1) \\ \hline
Ce(2)&6\textit{c} &0.338\newline(4)& 0 &	0 &	0.1414(1) &	12(1) \\ \hline
Mg(2)&6\textit{c} & 0.662\newline(4)& 0 &	0 &	0.1414(1)	& 12(1) \\ \hline
Co(1)&3\textit{b}&1& 0  &	0 &	$\frac{1}{2}$ &	11(1) \\ \hline
Co(2)&6\textit{c}&1& 0  &	0 &	0.3339(1) & 17(1) \\ \hline
Co(3)&18\textit{h}&1& 0.5014(1) & 0.4986(1) &	0.0840(1) &11(1) \\ \hline
\end{tabular}
\end{center}
\end{table}

\begin{figure}[!h]
\includegraphics[scale =0.35]{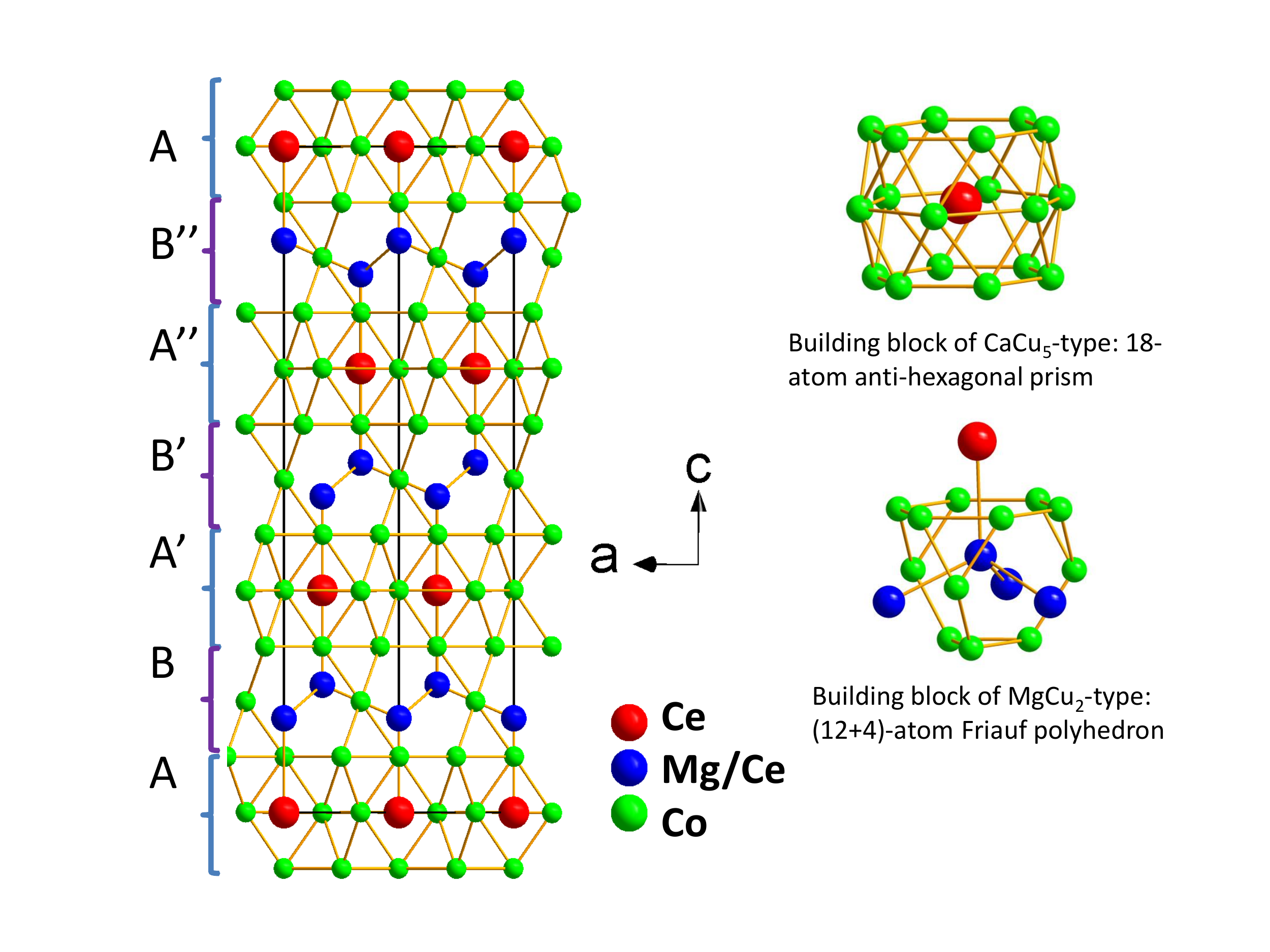}
\caption{The crystal structure for Ce$_{3-x}$Mg$_{x}$Co$_{9}$ showing the stacking sequence of CaCu$_5$ type plane (A) and MgCu$_2$ type plane (B) visualized along [010] direction.}
\label{Crystalstructure}
\end{figure}

A picture of as grown  Ce$_{1.66}$Mg$_{1.34}$Co$_{9}$ single crystals, a Laue backscattered photograph and single crystal surface diffracted monochromatic XRD data collected via Rigaku MiniFlex II diffractrometer with Bragg-Brentano geometry~\cite{Jesche2016PM}  are presented in pannels (a), (b) and (c) of FIG.~\ref{CombinedXRD} respectively. Both Laue and  monochromatic single crystalline XRD data confirmed that the single crystals grow with a planar morphology with the \textit{c}-axis perpendicular to the plates. 

\indent
The crystallographic data obtained from the single crystal X-ray diffraction for Ce$_{1.66}$Mg$_{1.34}$Co$_{9}$ grown out of Ce$_9$Mg$_{24}$Co$_{67}$ initial melt are summarized in TABLES.~\ref{tbl:crystaldata}, and ~\ref{tbl:atomcoord}. FIG.~\ref{CombinedXRD}(d) shows a powder X-ray diffraction pattern of the crushed single crystals which has some noticeable mismatch in observed and Rietveld refined intensity of \{00l\} families of the peaks, indicating a degree of  preferred orientation in the powder sample. As mentioned above, the relatively small crystal size made it difficult to acquire enough powder sample to obtain less noisy XRD data and better statistics in the Rietveld refinement. To reduce the intensity mismatch, a preferred orientation correction was employed in the Rietveld refinement using spherical harmonics up to 12$^{th}$ order and absorption correction for plate like grains in the powder sample. The Rietveld refined lattice parameters for powder XRD data of the single crystal Ce$_{1.66}$Mg$_{1.34}$Co$_{9}$ are \textit{a} = $4.923(1)~\AA$ and \textit{c} = $24.026(1)~\AA$ with \textit{R}$_{p}$ = 0.09 which are in close agreement (within 2-3$\sigma$) with single crystal XRD data as shown in TABLE \ref{tbl:crystaldata}. The single crystal XRD composition was Ce$_{13.92}$Mg$_{11.08}$Co$_{75}$ (Ce$_{1.662(4)}$Mg$_{1.338(4)}$Co$_{9}$). Although we did not make quantitative compositional analysis measurement on the single crystalline sample with EDS (The crystals were too thin to readily polish and small droplets of Mg rich flux was on their surfaces.) we could detect the minor presence of Ta, (up to 1 at.\%) most likely caused by slight dissolution of the inner wall surfaces of Ta reaction container and diffusion of Ta atoms into the reaction liquid during the long term dwelling process at the maximum temperature of $1200^\circ$~C as well as at ramping down to ~$1100^\circ$~C over 75 hours. However, an attempt to solve the crystal structure along with inclusion of Ta in any Wyckoff sites or interstitial sites was unsuccessful. We believe that Ta is distributed in our crystals in the form of nano-sized precipitates rather than incorporated into interstices of the crystals structure. 

\indent
The crystal structure of Ce$_{3-x}$Mg$_{x}$Co$_{9}$ is rhombohedral and belongs to  the PuNi$_3$ type structure\cite{Denys2012JSSC}. Similar to the R$_{3-x}$Mg$_x$Ni$_{9}$ series, the Co-containing structure is an intergrowth of CaCu$_{5}$-type (A) and MgCu$_{2}$-type (B) building blocks with a repeating sequence of ABA'B'A"B"A as shown in FIG.~\ref{Crystalstructure}. Here A', B' and A", B" are introduced to show the relative translation of the growth layers with respect to c-axis during stacking. There are two independent sites for Ce atoms in this structure: one 3\textit{a} site is located at the center of a face-shared anti-hexagonal prism defined by 18 Co atoms; the other 6\textit{c} site is surrounded by 12 Co atoms defining a truncated tetrahedron plus four capping atoms at longer distances. As expected from the polyhedra volume, the statistically distributed Ce/Mg mixtures prefer to occupy the Wyckoff 6\textit{c} site with its smaller volume.

\subsubsection{Polycrystalline samples: Composition and lattice parameters}

\begin{table}[h]
\begin{center}
\caption{The comparison of the loaded compositions with the EDS determined composition. A nominal presence of Ta (up to 1 at.\%) was found in the homogeneous Ce$_{3-x}$Mg$_{x}$Co$_{9}$ samples. Some of the higher-Mg samples  showed traces of a TaCo$_3$ impurity phase and low-Mg content samples showed a TaCo$_2$ phase.} 
\label{tbl:EDS}
\begin{tabular}{| p{3cm} | p{3cm} |p{2cm}|}
\hline
Loaded composition \newline (Nominal) & EDS composition& Rietveld Refinement \% of majority phase:\newline Ce$_{3-x}$Mg$_{x}$Co$_{9}$  \\ \hline
CeCo$_3$ \newline (arcmelted and annealed at $900^\circ$C for 7 days)&CeCo$_3$ + CeCo$_2$&$\geq88\%$\\ \hline
Ce$_{2.75}$Mg$_{0.25}$Co$_{9}$ & Ce$_{2.82}$Mg$_{0.18}$Co$_{9}$ + Ce$_{0.86}$Mg$_{0.14}$Co$_{2}$ + TaCo$_2$ &$\geq67\%$\\ \hline
Ce$_{2.50}$Mg$_{0.5}$Co$_{9}$ & Ce$_{2.66}$Mg$_{0.34}$Co$_{9}$ + Ce$_{0.77}$Mg$_{0.23}$Co$_{2}$ & $\geq76\%$\\ \hline
Ce$_{2.25}$Mg$_{0.75}$Co$_{9}$ &  Ce$_{2.40}$Mg$_{0.60}$Co$_{9}$ + Ce$_{0.80}$Mg$_{0.20}$Co$_{2}$ & $\geq80\%$  \\ \hline
Ce$_{2}$Mg$_{1.0}$Co$_{9}$ & Ce$_{2.18}$Mg$_{0.82}$Co$_{9}$ & $\approx$ single phase \\ \hline
Ce$_{1.67}$Mg$_{1.33}$Co$_{9}$ &Ce$_{1.89}$Mg$_{1.11}$Co$_{9}$ &  $\approx$ single phase\\ \hline
Ce$_{1.5}$Mg$_{1.5}$Co$_{9}$ &  Ce$_{1.77}$Mg$_{1.23}$Co$_{9}$& $\approx$ single phase\\ \hline
Ce$_{1.33}$Mg$_{1.67}$Co$_{9}$ & Ce$_{1.65}$Mg$_{1.35}$Co$_{9}$&  $\approx$ single phase\\ \hline
CeMg$_{2}$Co$_{9}$ & Ce$_{1.42}$Mg$_{1.58}$Co$_{9}$ + MgCo$_2$ + Co& $\geq50\%$\\ \hline
\end{tabular}
\end{center}
\end{table}

\begin{figure}[!h]
\includegraphics[scale =0.6]{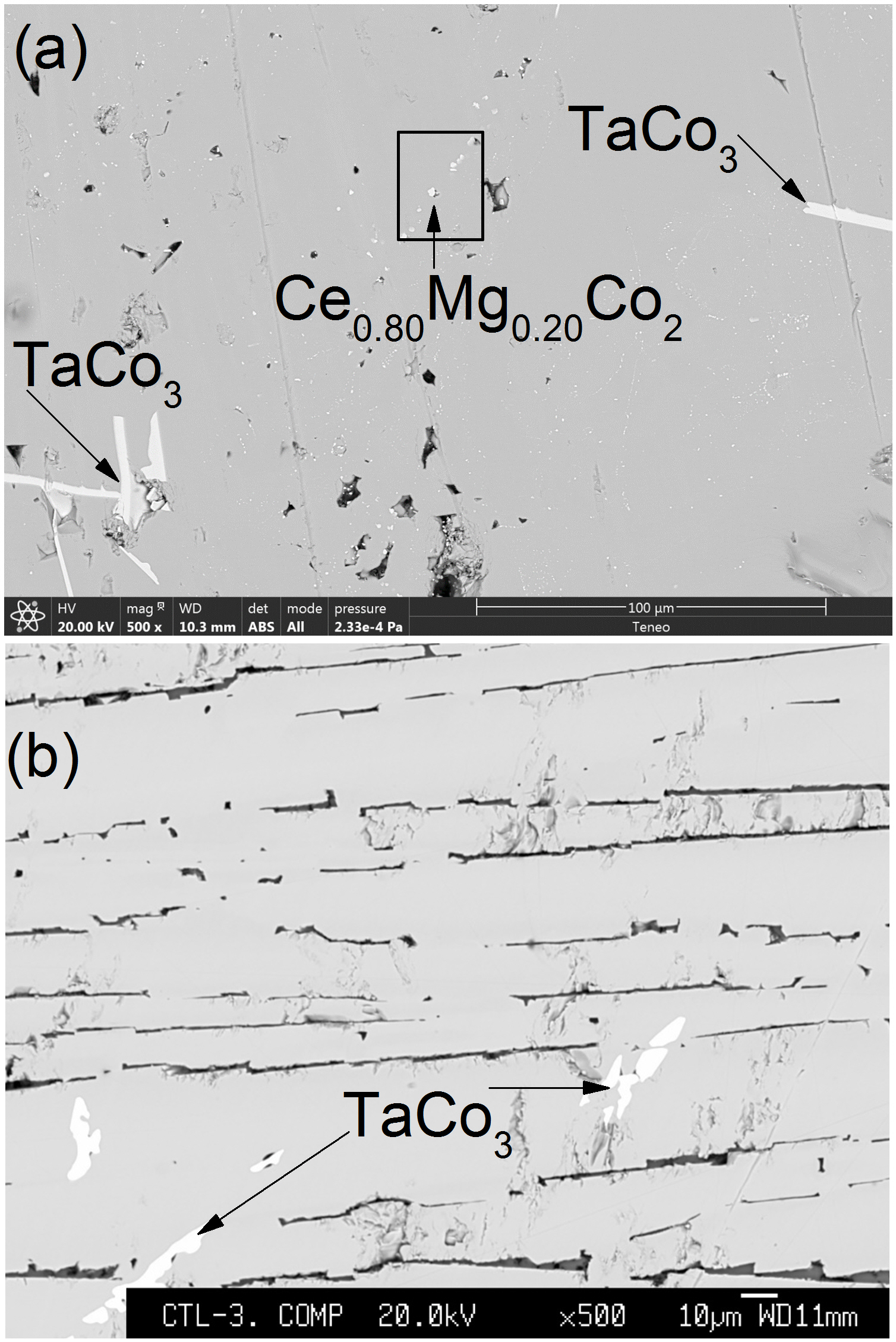}
\caption{SEM images of mixed phase and predominantly single phase Ce$_{3-x}$Mg$_{x}$Co$_{9}$ samples (a) Nominal Ce$_{2.25}$Mg$_{0.75}$Co$_{9}$ which gave a mixture of Ce$_{2.40}$Mg$_{0.60}$Co$_{9}$ (majority phase) and Ce$_{0.80}$Mg$_{0.20}$Co$_{2}$ (minority phase demonstrated as faint small white spots along the diagonal of the black box in the panel (a)) and large white stripes of TaCo$_3$ impurity phase  (b) Predominantly single phase Ce$_{1.89}$Mg$_{1.11}$Co$_{9}$ along with some traces of TaCo$_3$ impurity. The black parallel grooves in the image represent the cracks in the polished sample.}
\label{mixedandsingle}
\end{figure}

\begin{figure}[!h]
\includegraphics[scale =0.35]{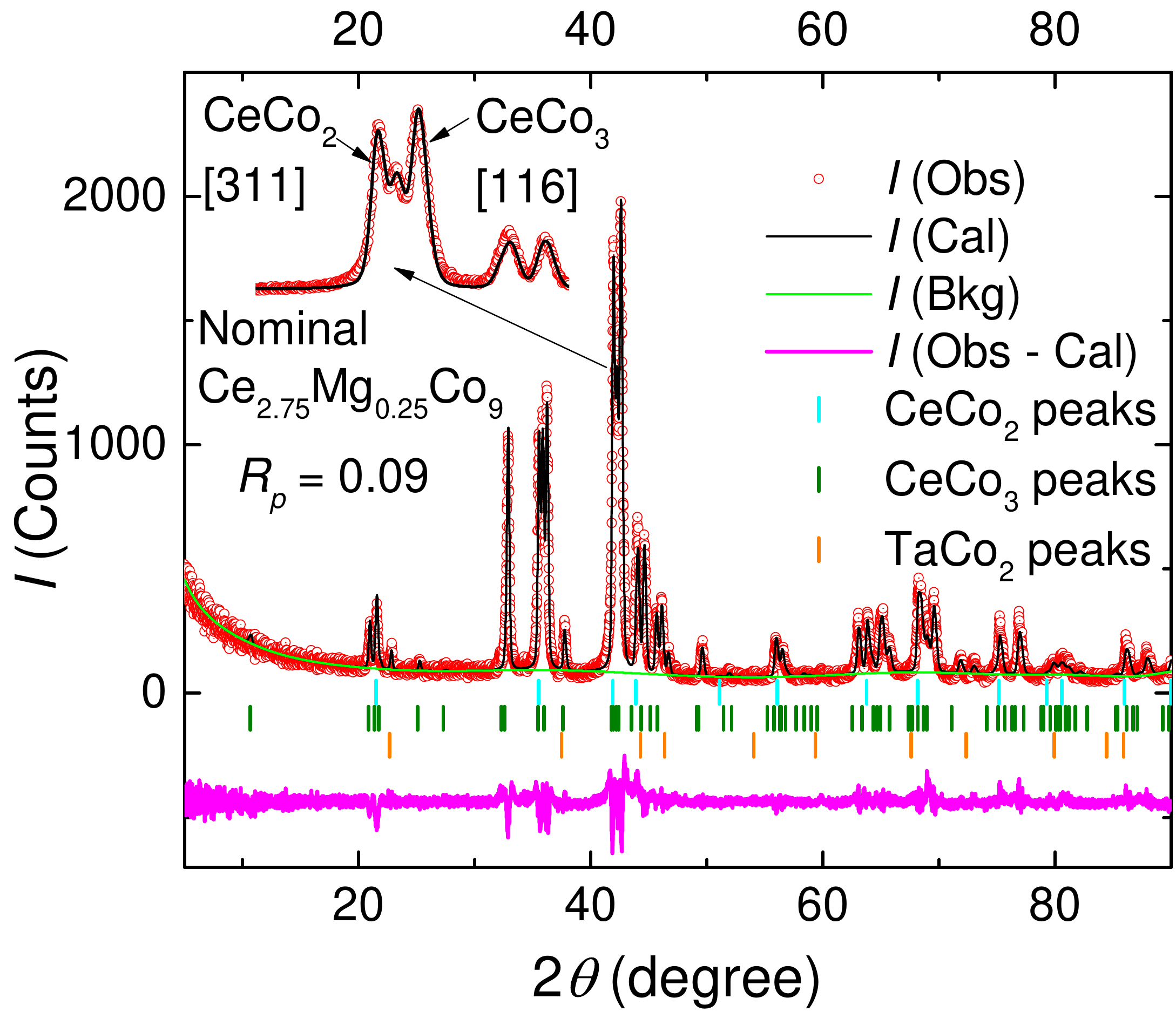}
\caption{A typical example of multiphase polycrystalline XRD pattern for  nominal Ce$_{2.75}$Mg$_{0.25}$Co$_{9}$ sample. The enlarged peak on the left top of the graph shows the broadening of the highest intensity peak of Ce$_{3-x}$Mg$_{x}$Co$_{9}$ around $2\theta$ value of $42^\circ$ due to presence of Mg doped CeCo$_2$ diffraction peak. \textit{I}(Obs), \textit{I}(Cal) and \textit{I}(Bkg) are experimental, Rietveld refined and instrumental background data respectively. The lower section of the graph shows the Bragg's peaks positions with different coloured vertical lines for phases shown in the graph and the differential X-ray diffractrogram \textit{I}(Obs-Cal).}
\label{TL791-Bn}
\end{figure}

\begin{figure}[!h]
\includegraphics[scale =0.35]{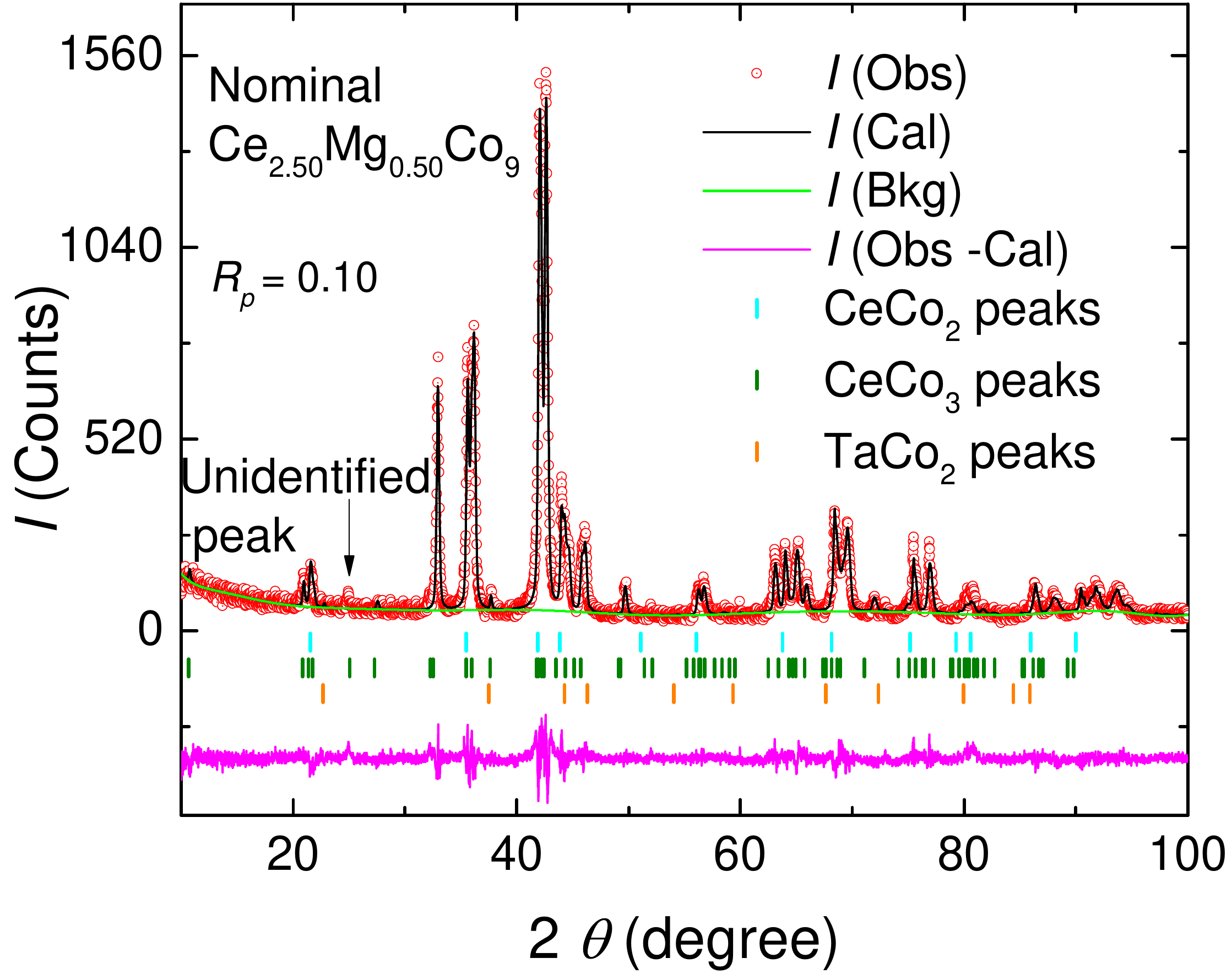}
\caption{Powder XRD pattern for  nominal Ce$_{2.50}$Mg$_{0.50}$Co$_{9}$ sample. TaCo$_2$ phase is almost not detectable for XRD however an unidentified XRD peak is observed around 2$\theta$ value of $25^\circ$.  \textit{I}(Obs), \textit{I}(Cal) and \textit{I}(Bkg) are experimental, Rietveld refined and instrumental background data respectively. The lower section of the graph shows the Bragg's peaks positions with different coloured vertical lines for phases shown in the graph and the differential X-ray diffractrogram \textit{I}(Obs-Cal)}
\label{TL801Mg0p5}
\end{figure}

\begin{figure}[!h]
\includegraphics[scale =0.35]{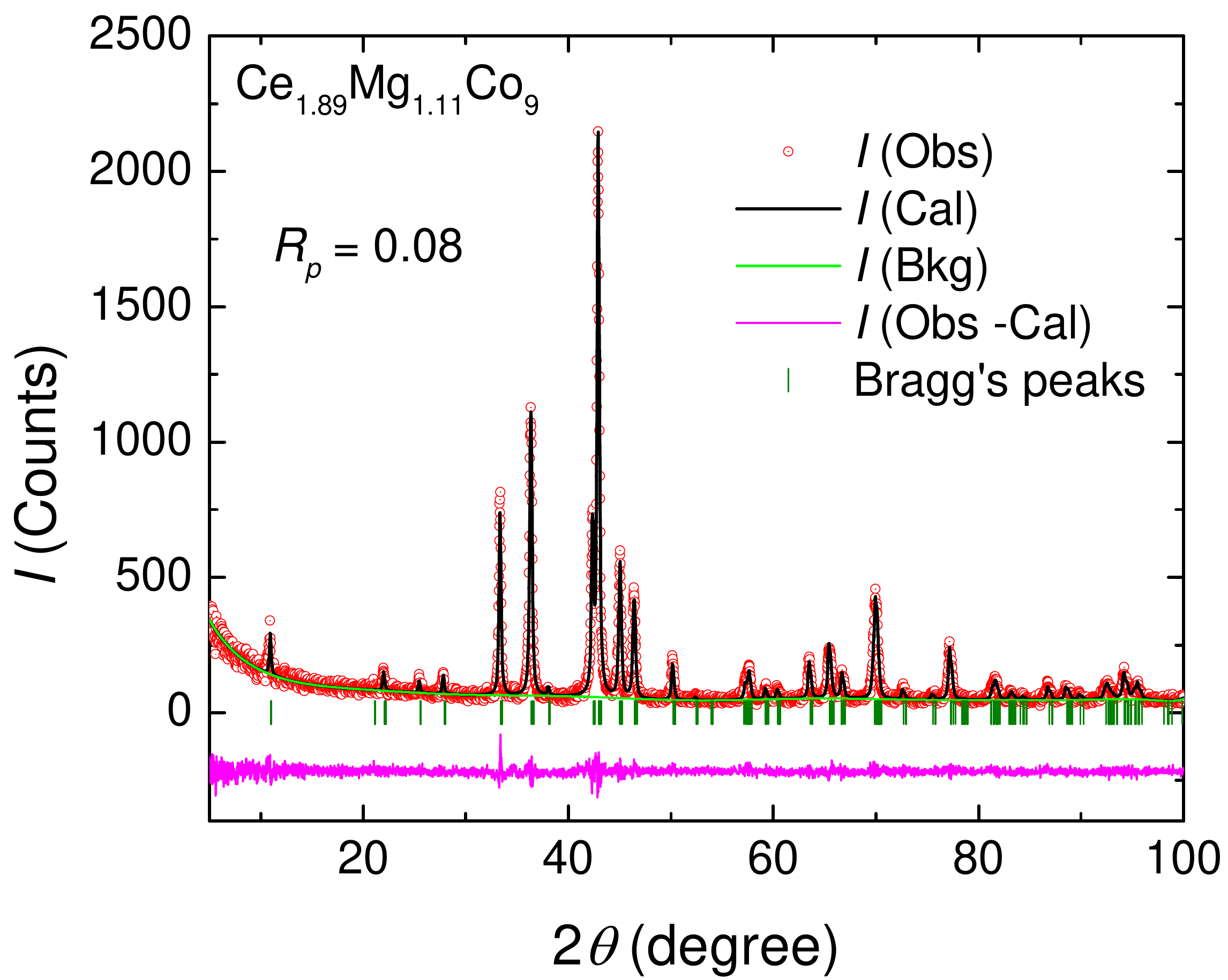}
\caption{A typical example of predominantly single phase polycrystalline XRD pattern for EDS characterized  Ce$_{1.89}$Mg$_{1.11}$Co$_{9}$ sample. \textit{I}(Obs), \textit{I}(Cal) and \textit{I}(Bkg) are experimental, Rietveld refined and instrumental background data respectively. The lower section of the graph shows the Bragg's peaks positions with olive vertical lines and the differential X-ray diffractrogram \textit{I}(Obs-Cal)}
\label{TL501}
\end{figure}

\begin{figure}[!htb]
\includegraphics[scale =0.35]{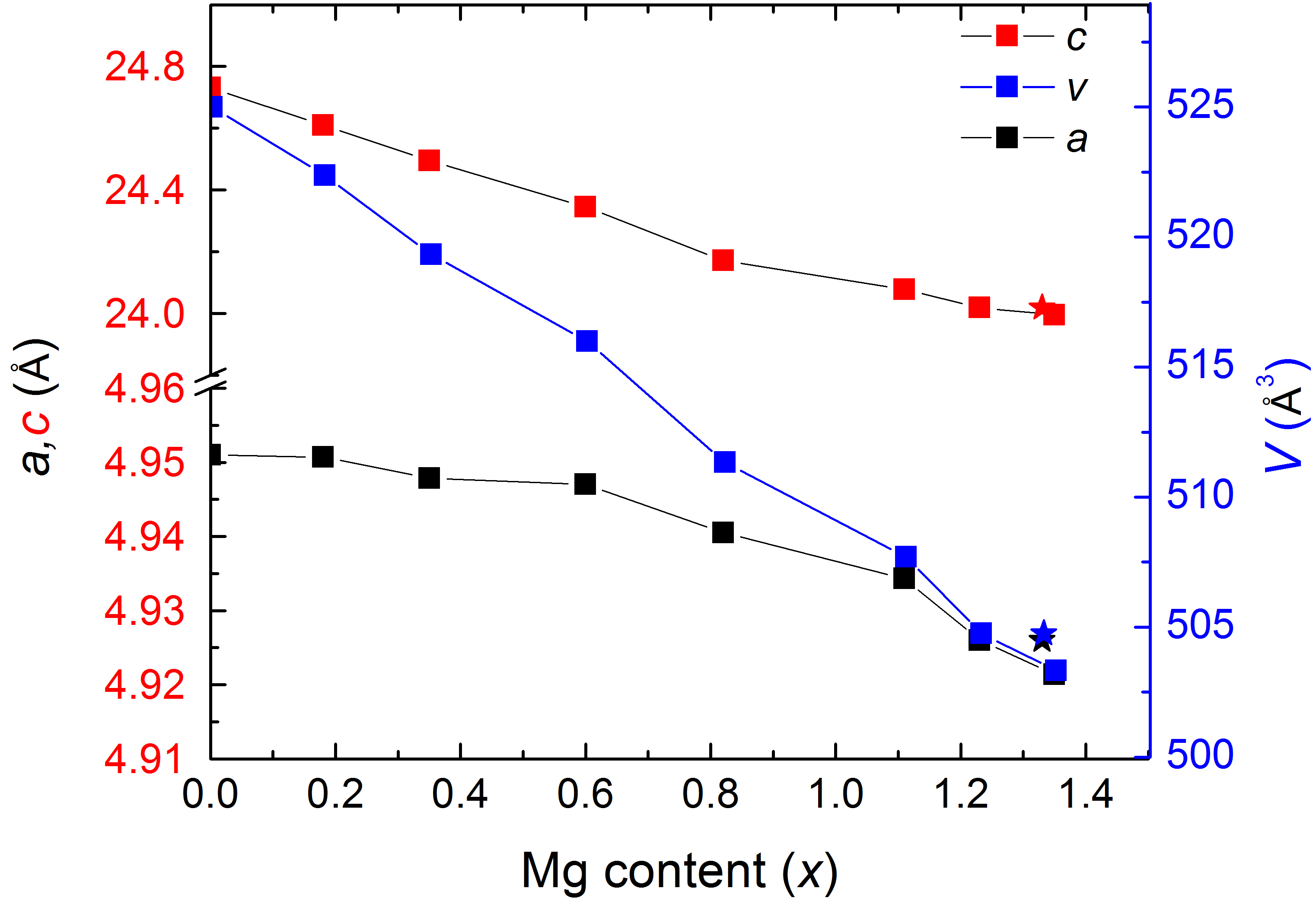}
\caption{Variation of lattice parameters (\textit{a},\textit{c}) and unit cell volume (\textit{v}) of polycrystalline  with Mg content inferred from EDS. Cubic Ce$_{1-x}$Mg$_{x}$Co$_{2}$ types impurity phases were obtained for $x\leq 0.6$ and predominant single phase Ce$_{3-x}$Mg$_{x}$Co$_{9}$ is obtained for $0.6 < x \leq 1.4$.   The lattice parameters for single crystalline  Ce$_{1.66}$Mg$_{1.34}$Co$_9$ are presented with corresponding color stars. The uncertainty in the refined lattice parameters is less than 0.01\% of the reported lattice parameters and too small to clearly show as an error bar in the diagram. }
\label{a-c-V-f}
\end{figure}

The nominal and EDS composition of the  polycrystalline samples are presented in TABLE~\ref{tbl:EDS} along with the Rietveld refined percentage of the majority phase in the sample. The SEM images for mixed and predominantly single phase  Ce$_{3-x}$Mg$_{x}$Co$_{9}$ samples are presented in FIG.~\ref{mixedandsingle}. Panel (a) in FIG.~\ref{mixedandsingle} shows the mixed phase sample that forms out of a nominal composition Ce$_{2.25}$Mg$_{0.75}$Co$_9$ and the panel (b) shows the predominantly single phase sample with EDS composition Ce$_{1.89}$Mg$_{1.11}$Co$_9$.  In addition TABLE~\ref{tbl:EDS} summarizes phase analysis based on powder XRD data.
\indent

% For nominal $x<1$, mixtures of Ce$_{3-x}$Mg$_x$Co$_9$, Ce$_{1-x}$Mg$_x$Co$_2$ and TaCo$_2$ phases are present. At lowest finite Mg content (nominal \textit{x} = 0.25) sample, significant amount of TaCo$_2$ was detected. TaCo$_2$ concentration was significantly reduced in nominal Mg content  \textit{x} = 0.5 sample and only trace of TaCo$_3$ was found in higher Mg containing samples (nominal x$\geq0.75$). For $1\leq x\leq 1.67$, the Ce$_{3-x}$Mg$_x$Co$_9$ phase crystallizes as a predominant single phase. From the analysis of the majority phase \% in the TABLE~\ref{tbl:EDS}, it might be concluded that the annealing of Ce$_{3-x}$Mg$_x$Co$_9$ sample is better effective with finite amount of the Mg in the initial composition. 

\indent

The crystallographic information file obtained from  single crystal XRD was used to perform Rietveld refinement of the powder XRD data of polycrystal samples listed in TABLE~\ref{tbl:EDS}. Rietveld refined XRD patterns for a multiple phase polycrystalline samples (Nominal Ce$_{2.75}$Mg$_{0.25}$Co$_9$ with $R_p$ = 0.09  and Ce$_{2.50}$Mg$_{0.50}$Co$_9$  with $R_p$ = 0.10)  and single phase polycrystalline  Ce$_{1.89}$Mg$_{1.11}$Co$_9$ (EDS composition) with $R_p$ = 0.08 are presented in FIG.~\ref{TL791-Bn},~\ref{TL801Mg0p5} and \ref{TL501} respectively. The melt-annealed nominal Ce$_{2.75}$Mg$_{0.25}$Co$_9$ sample contains $\sim 67\%$ of Ce$_{2.82}$Mg$_{0.18}$Co$_9$ phase, $\sim 23\%$ of Ce$_{0.86}$Mg$_{0.14}$Co$_2$ and $\sim 10\%$ of TaCo$_2$ inferred from Rietveld refinement. The TaCo$_2$ phase is not observed in the nominal Ce$_{2.50}$Mg$_{0.5}$Co$_9$ and higher content of Mg as shown in FIG.~\ref{TL801Mg0p5} and~\ref{TL501}. An unidentified XRD peak was observed in nominal Ce$_{2.50}$Mg$_{0.50}$Co$_9$  sample as shown in FIG.~\ref{TL801Mg0p5} however TaCo$_2$ phase was almost reduced to zero in comparison to nominal composition Ce$_{2.75}$Mg$_{0.25}$Co$_9$ as shown in FIG.~\ref{TL791-Bn}.  But traces of TaCo$_3$ phase was observed in predominantly single phases Ce$_{3-x}$Mg$_{x}$Co$_{9}$ samples. These results combined with the fact that even pure (Mg = 0) CeCo$_3$ remains mixed phase after 7 days of annealing, suggest that Mg be assisting the annealing of polycrystalline Ce$_{3-x}$Mg$_{x}$Co$_{9}$ samples.

\indent
 For nominal $x=2$ and higher, Ce$_{3-x}$Mg$_x$Co$_9$ can no longer be considered a clear majority phase with the presence of the significant amount of CoMg$_2$ and Co. Looking at the composition of the Ce$_{3-x}$Mg$_x$Co$_9$ alloys from EDS, it seems that $x\approx1.4$ is the maximum solid solubility. In the Nd$_{3-x}$Mg$_x$Co$_9$ alloys, the structure changes from the trigonal structure for $x\leq 1.5$ to a tetragonal structure at $x=2$ (YIn$_2$Ni$_9$-type)~\cite{Shtender2017JAC}. The solubility range of Mg in CeCo$_3$ is therefore similar but we do not observe a phase corresponding with the YIn$_2$Ni$_9$-type structure for $x\geq 2$. Instead, a three phase region of  MgCo$_2$, Ce$_{3-x}$Mg$_x$Co$_9$ and Co is observed (see TABLE~\ref{tbl:EDS}). The compositional range of our Ce$_{3-x}$Mg$_x$Co$_9$ samples is summarized in FIG.~\ref{Composition}. 

\indent
The variation of the polycrystalline  lattice parameters and unit cell volume as a function of the Mg content in the Ce$_{3-x}$Mg$_x$Co$_9$ phase as determined from EDS is shown in FIG.~\ref{a-c-V-f}. As expected, the substitution of Ce by Mg results in the reduction of the unit cell volume, similar to the case of Nd$_{3-x}$Mg$_x$Co$_9$ alloys~\cite{Shtender2017JAC}.  Neither the \textit{a} or \textit{c}-lattice parameter follows a linear relation with Mg content \textit{x}. The variation in \textit{a} lattice parameter shows the slight positive deviation and \textit{c} lattice parameter shows the slight negative deviation starting in the middle of the single phase region. The negative deviation of lattice parameter \textit{c} might indicate that the covalent bonding is increased along that direction. 
 It should be noted that the lattice parameters (\textit{a}, \textit{c}, \textit{v}) and composition inferred from the single crystal X-rays (shown as corresponding color $\star$s ) agree very well with what we inferred from EDS measurements on the polycrystalline samples.

\section{Magnetic Properties}
\begin{figure}[!htb]
\includegraphics[scale =0.35]{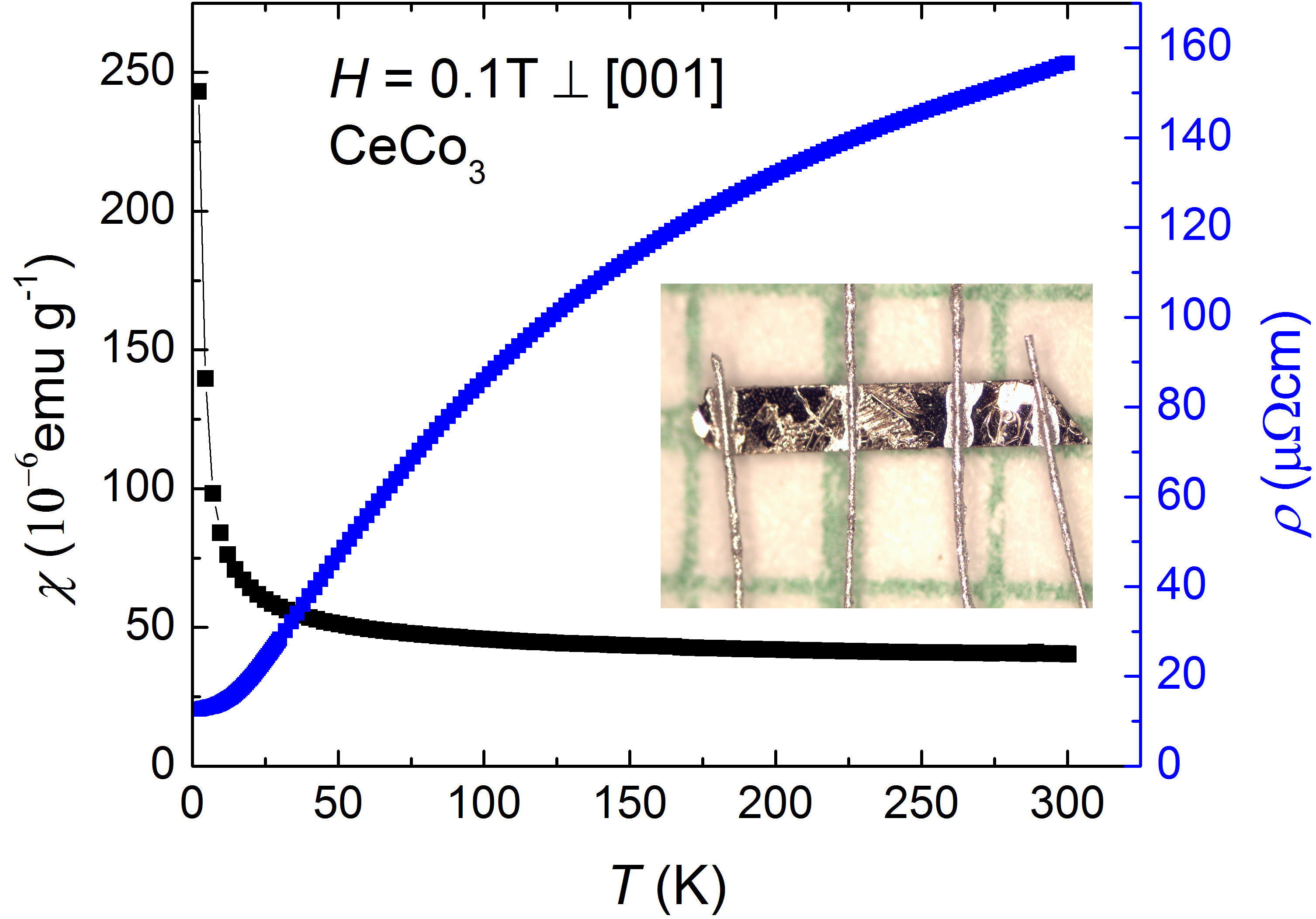}
\caption{Temperature dependent susceptibility (\textit{H} = 0.1 T $\perp$ [001] ) and electrical resistivity (excitation current $\perp$ to [001]) of CeCo$_{3}$ single crystal. The picture in the inset shows the resistivity bar for four probe measurement. The sample was 60 micrometer thick.}
\label{CeCo3resistivity}
\end{figure}

Previously reported data do not agree about the magnetic properties of the parent compound, CeCo$_{3}$. Lemaire reported CeCo$_3$ as a ferromagnetic material with Curie temperature $78$~K \cite{LEMAIRCobalt}. Buschow identified it as a Pauli-paramagnetic phase however he left a room for further investigation mentioning CeCo$_3$ could be ferromagnetic below $10$~K\cite{Buschow1980JLCM}. To clarify this issue, we measured the temperature dependent magnetization and  electric resistivity of our single crystalline samples down to $2$~K as shown in FIG.~\ref{CeCo3resistivity}. The magnetization data shows no signature of a phase transition and is only weakly temperature dependent and is consistent with a Pauli-paramagnet and an impurity tail below $20$~K. Assuming Curie tail is because of Ce$^{3+}$ magnetic ions in the single crystalline CeCo$_3$ sample, magnetic susceptibility was fitted to Curie-Weiss law up to $150$~K as:\begin{center}
$\chi(T) = \chi_0 + \frac{C}{T - \theta}$
\end{center} where $\chi_0$ is high temperature asymptotic susceptibility, \textit{C} is Curie constant and $\theta$ is Curie-Weiss temperature. 
The concentration of Ce$^{3+}$ ions was estimated to be $\sim 20$\% ( with $\theta$ = $3.8$~K) using the spin only moment of $2.54\mu_B$ per Ce$^{3+}$ ion. 
 The electrical resistivity  does not show any signature of a loss of spin disorder scattering that would be anticipated for a magnetic phase transition. 
    
\begin{figure}[!htb]
\includegraphics[scale =0.35]{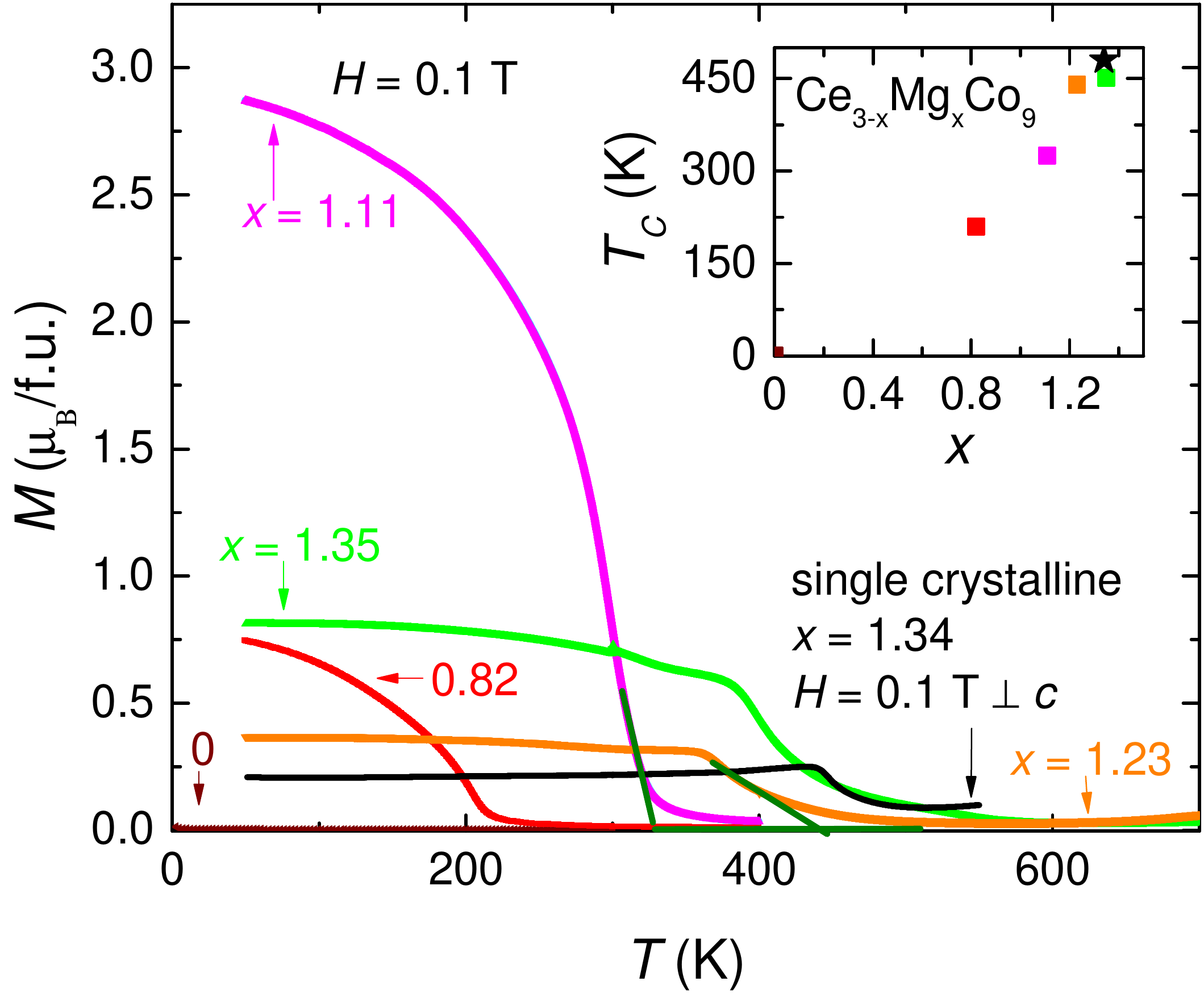}
\caption{Temperature dependent magnetization of single crystalline (\textit{x} = 0 and 1.34)  polycrystalline Ce$_{3-x}$Mg$_x$Co$_9$ measured under a magnetic field of $0.1$~T. The arrow pointed Mg content (\textit{x}) for each \textit{M(T)} graph are inferred from EDS analysis. The olive coloured straight lines above and below the point of inflection of \textit{M(T)} data for for \textit{x} = 1.11 and \textit{x} = 1.23 shows the scheme for inferring the Curie temperature. Thus obtained Curie temperatures and Mg content phase diagram is presented in the inset. The black $\star$ represents the Curie temperature inferred from the tangents intersection scheme for single crystalline Ce$_{1.66}$Mg$_{1.34}$Co$_{9}$ sample on the \textit{M(T)} data measured along the hard axis of the plate.}
\label{MT-Poly}
\end{figure}

The temperature dependence of the magnetization data of the single phase  Ce$_{3-x}$Mg$_x$Co$_9$ polycrystalline samples are shown in FIG.~\ref{MT-Poly}. A rapid increase of the magnetization upon cooling below the Curie temperature is observed  for \textit{x} = 0.82 - 1.35 indicating the appearance of ferromagnetism with Mg substitution.  

\indent
 The Curie temperature $T_\textrm{C}$ is estimated as the cross-point of linear extrapolations of two tangents to magnetization curve around the point of inflection as indicated on the curve $x=1.11$ and \textit{x} = 1.23 in FIG.~\ref{MT-Poly}. The variation of $T_\textrm{C}$ with Mg substitution is shown in the inset. The Curie temperature increases with Mg concentration and reaches as high as $450$\,K for $x=1.35$. 
 
\indent
% Such a transformation of Pauli-paramagnetic CeCo$_3$ alongwith large increase in $T_\textrm{C}$ is rather surprising, especially when substituting Ce by Mg. Given that CeCo$_3$ is a reported Pauli paramagnet~\cite{Buschow1980JLCM}, Ce atoms are non-magnetic in the structure. Interestingly, the formation of the hydride CeCo$_3$H$_4$ induces an expansion of the lattice and the appearance of ferromagnetism~\cite{Buschow1980JLCM}, whereas in the case of Mg substitution, the lattice is compressed. Although the experimental evidences are not enough to explain the reason of induction of  ferromagnetism in Ce$_{3-x}$Mg$_x$Co$_9$, further experiments or theoretical investigations focused on band-filling effect and tuning of Columb interaction to meet the Stoner enhancement criterion could be interesting. 

The substitution of Mg for Ce changes a Pauli-paramagnet (CeCo$_3$) into a ferromagnet that has T$_C$ increase with Mg content. The most likely reason for this is associated with valency and band filling. In CeCo$_3$, the Ce is essentially non-moment bearing; this implies a Ce$^{4+}$ valency. As Mg$^{2+}$ is added, there will be a clear change in band filling that most likely leads to Stoner-type magnetism associated with the Co 3-d bands. Further work, both computational and experimental will be needed to better appreciate the origin of the observed ferromagnetism.
   
\begin{figure}[!htb]
\includegraphics[scale =0.35]{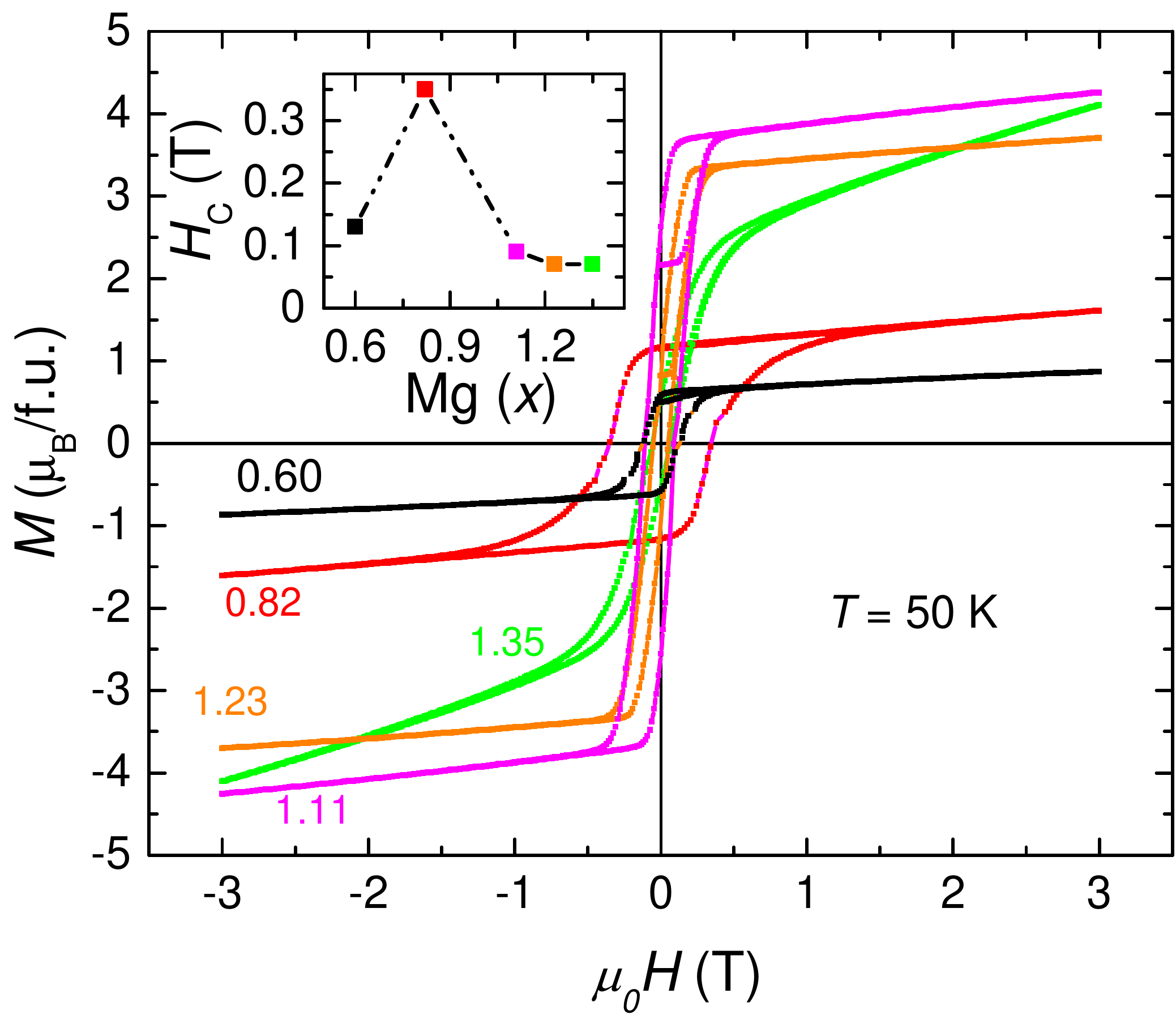}
\caption{Magnetic hysteresis loop of various annealed polycrystalline Ce$_{3-x}$Mg$_x$Co$_9$ samples at $50$~K. The values of Mg content are from EDS analysis. The inset shows the variation of the observed coercivity fields of the samples as a function of Mg content.}
\label{MHpoly}
\end{figure}

Figure~\ref{MHpoly} shows the magnetic hysteresis curves of various polycrystals along with observed coercivity fields at $50$~K (inset). The coercivity field increases with Mg content, becomes maximum ($\sim0.35$~T) for \textit{x} = 0.82 and then decreases. The observation of coercivity is consistent with the axial nature of the magnetic anisotropy as will be detailed below. The abruptly increased coercivity for \textit{ x} = 0.82  and spontaneous magnetization for \textit{ x} = 1.11 magnesium containing phases were reproduced in multiple samples.  It should be noted that differing saturation values for differing \textit{x} are most likely due to non-random distribution of grains in these as-cast samples. In addition, the non-saturating behaviour of the \textit{x} = 1.35 sample indicates either the presence of the preferred orientations of the grains with their hard axis along the applied field or presence of some anisotropic magnetic impurity in the sample. The former argument can be easily visualized in \textit{M(T}) data as \textit{x} = 1.35 polycrystalline sample and single crystalline Ce$_{1.66}$Mg$_{1.34}$Co$_9$ \textit{M(T)} data along the hard axis are almost identical in nature as shown in FIG.~\ref{MT-Poly}. The fact that as-cast samples show coercivity is promising for the development of permanent magnets out of this system.

\begin{figure}
\includegraphics[scale =0.35]{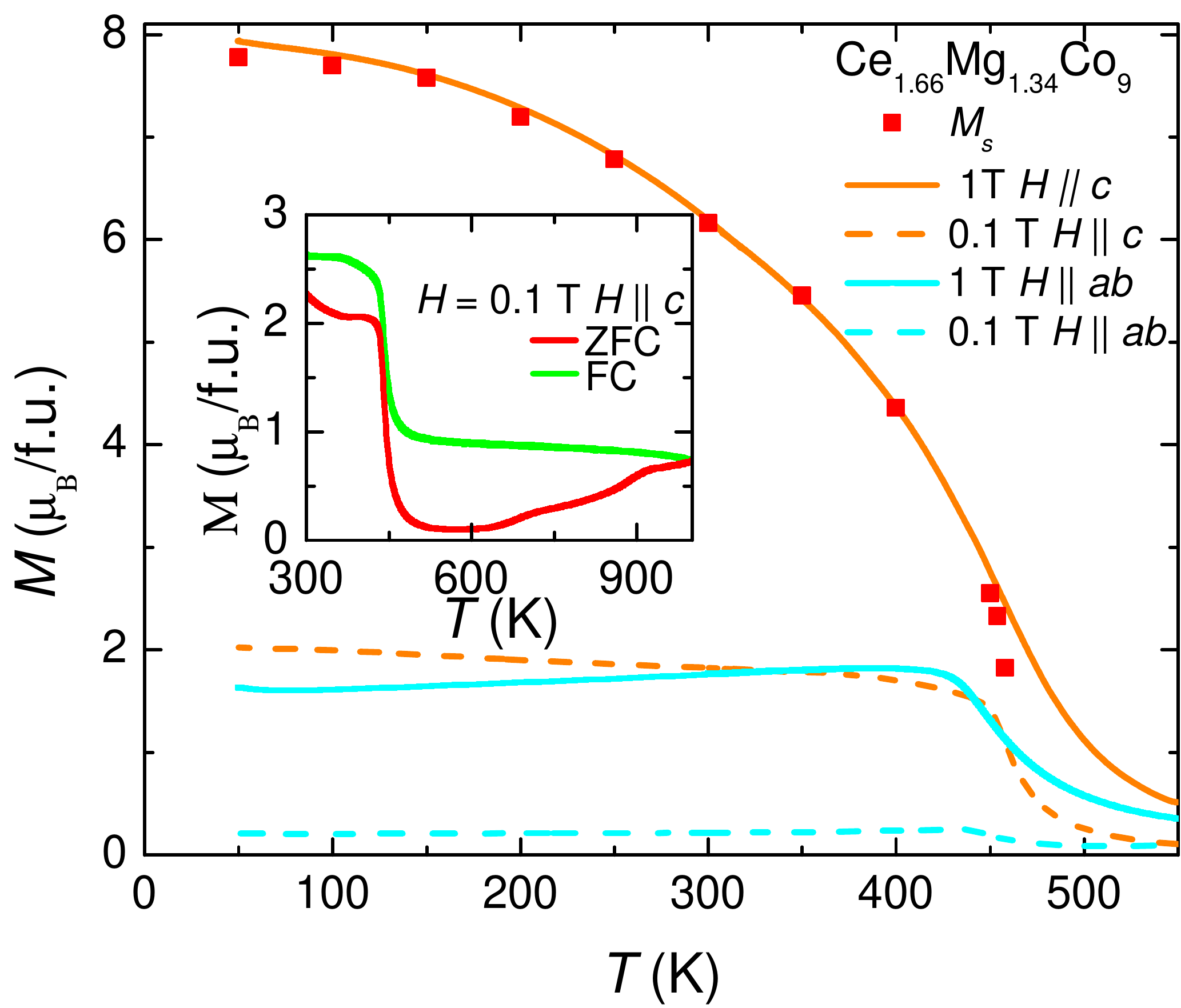}
\caption{Anisotropic temperature dependent magnetization for Ce$_{1.66}$Mg$_{1.34}$Co$_9$ at various applied fields and directions. Dashed lines are for \textit{H} = 0.1 T an solid lines are for \textit{H} = 1.0 T. Individual data points (red squares) are spontaneous magnetization,\textit{ M}$_S$ inferred from \textit{M(H)} isotherms. The inset shows the zero field cooled (ZFC) and field cooled (FC) magnetization data along the easy axis of magnetization to temperature above the sample decomposition point.}
\label{combinedMsMt}
\end{figure}

\indent
Our single crystal sample can provide further insight into this system's promise as a permanent magnet material. FIG.~\ref{combinedMsMt} shows the temperature dependence of the magnetization parallel and perpendicular to the c-axis of the single crystal of Ce$_{1.66}$Mg$_{1.34}$Co$_9$ up to $550$~K. The c-axis is the easy axis of magnetization and the saturation magnetization at low temperature is $8$~$\mu_B$/f.u. The spontaneous magnetization data points for  Ce$_{1.66}$Mg$_{1.34}$Co$_9$, represented with red squares, are obtained by the Y-intercept of linear fitting of the high field \textit{M(H)} data ($>$1.5~T) in the first quadrant. No hysteresis was observed in the \textit{M(H)} loops. The absence of hysteresis in \textit{M(H)} isotherms measured in single crystals is due to absence of grain boundaries, defects and impurity atoms capable of pinning of the ferromagnetic domains. When measurements were performed up to $1000$~K, we noticed a non-reproducibility of the results as shown in the inset of FIG.~\ref{combinedMsMt}. Thermo-gravimetric analysis (not shown here) showed evidence for a decomposition of the samples, this degradation was not observed when limiting the measurements to a maximum temperature of $550$~K.

\begin{figure}
\includegraphics[scale =0.35]{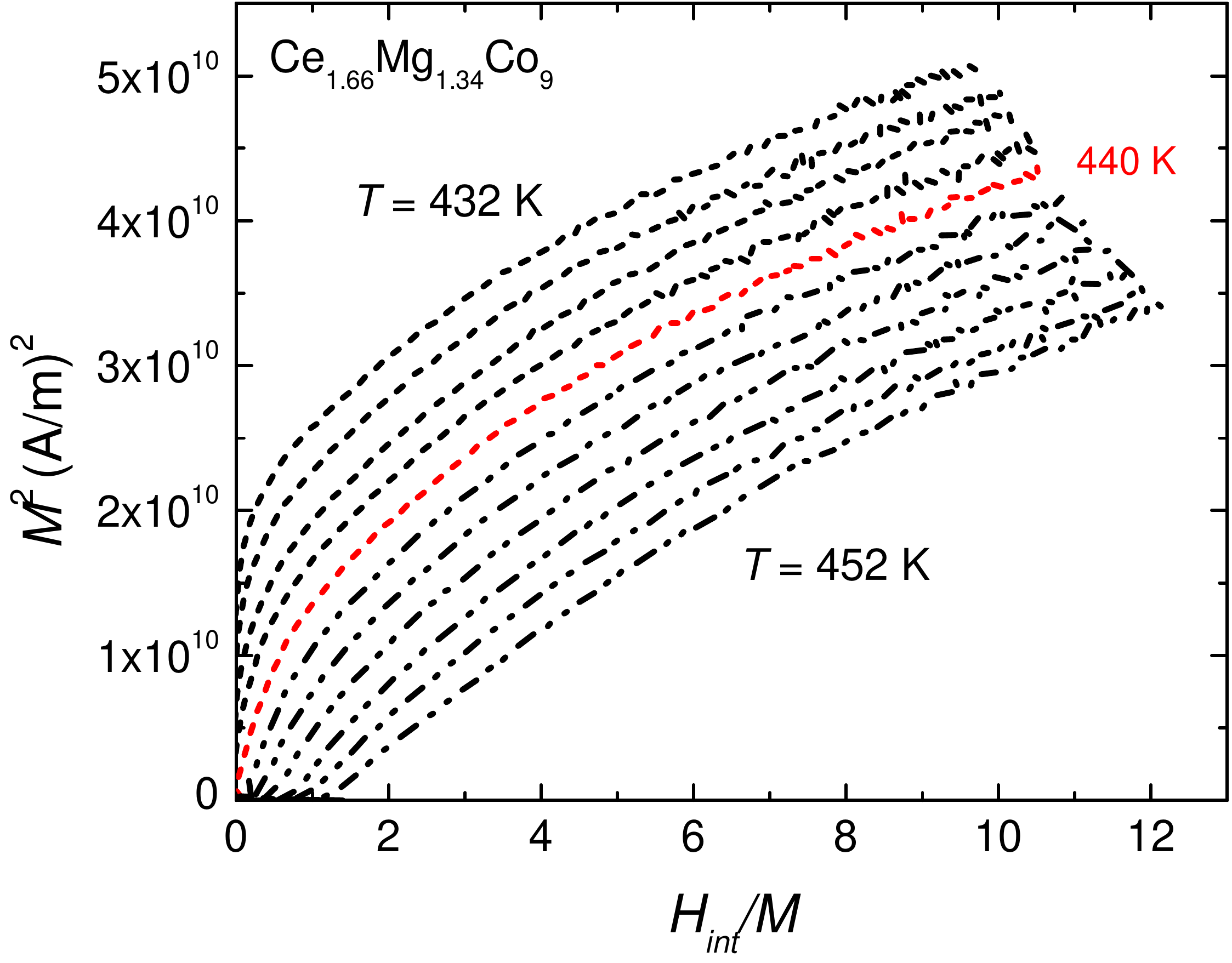}
\caption{Arrott plot for a Ce$_{1.66}$Mg$_{1.34}$Co$_9$ single crystal within the temperature range of $432$~K to $452$~K at a step of 2 K between adjacent curves. The Curie temperature is determined to be $440$~K.}
\label{Arrott}
\end{figure}

An Arrott plot with several isotherms near the Curie temperature is shown in FIG.~\ref{Arrott}. The internal magnetic field $H_{int}$ was determined using the relation $H_{int}= H_{app}-N*M$, where $H_{app}$ is the applied field, $N$ is the demagnetization factor which is experimentally determined along the easy axis ($N_c = 0.78$)~\cite{Lamichhane2015,HfZrMnPTej} and $M$ is the magnetization. The Arrott curves are not linear indicating Ce$_{1.66}$Mg$_{1.34}$Co$_9$ does not follow the mean field theory. The Curie temperature of Ce$_{1.66}$Mg$_{1.34}$Co$_9$ is $440$~K since Arrott curve corresponding to $440$~K passes through the origin. This is comparable with the values of T$_C$ obtained from polycrystals of similar composition (see inset of FIG.~\ref{MT-Poly}).

\begin{figure}[!h]
\includegraphics[scale =0.35]{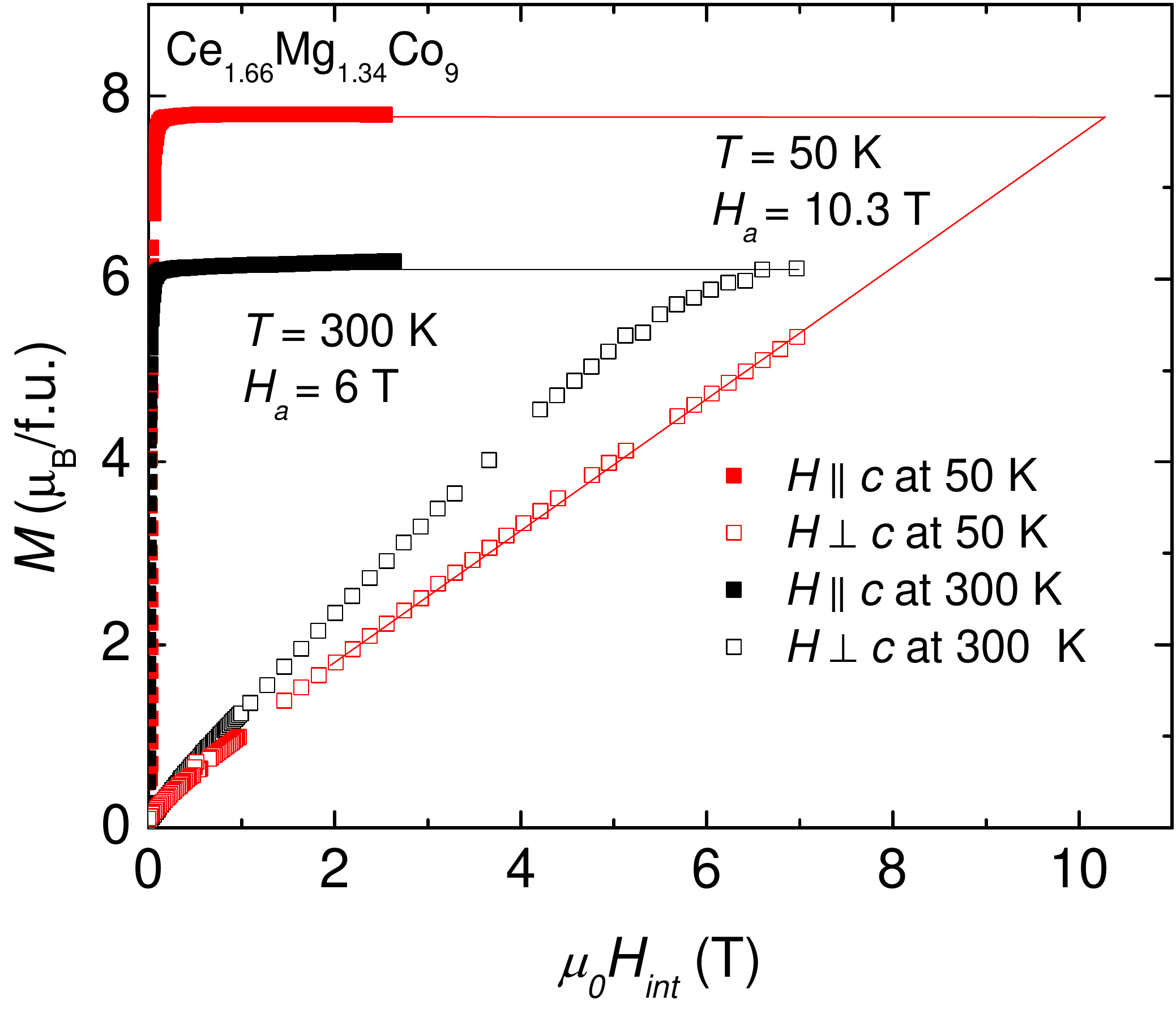}
\caption{Anisotropic field dependent magnetization of Ce$_{1.66}$Mg$_{1.34}$Co$_9$ at $50$~K (\textit{H}$_{a}\approx$10$~T)$ and $300$~K (\textit{H}$_{a}\approx6$~T).}
\label{Anisotropyfieldfu}
\end{figure}

\begin{figure}
\includegraphics[scale =0.35]{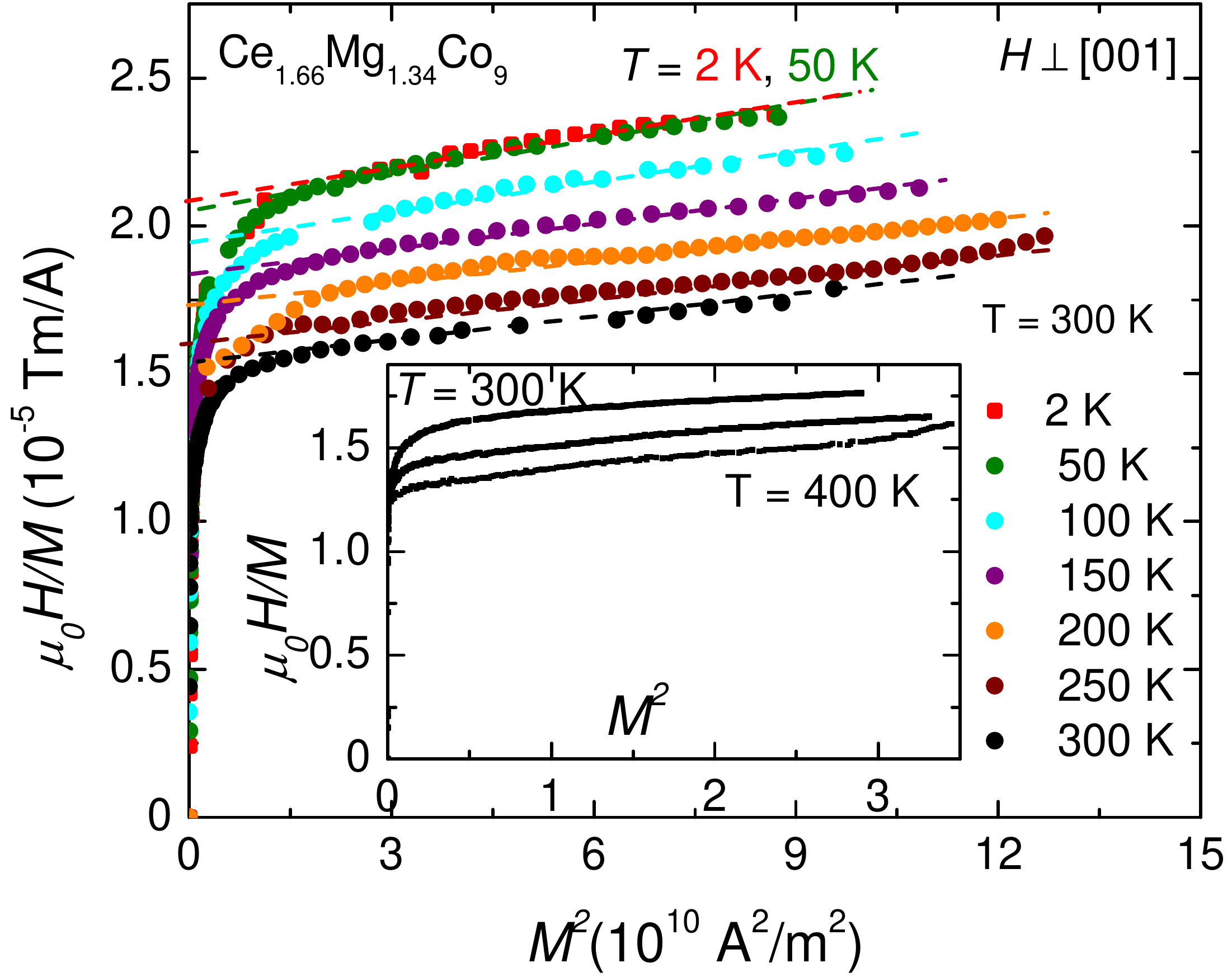}
\caption{Sucksmith-Thompson plot for Ce$_{1.66}$Mg$_{1.34}$Co$_{9}$ to obtain anisotropy constants \textit{K}$_{1}$ and \textit{K}$_{2}$. The field dependent magnetization data were measured up to $7$~T along the hard axis of magnetization. The dotted straight lines are the linear fit to $\frac{\mu_0 H}{M}$ at high field magnetization data to obtain the intercepts and slopes of the isotherms. The inset shows the Sucksmith-Thompson plots for VSM data measured along hard axis ( $\textit{H}\perp [001]$) up to $3$~T field.  \textit{K}$_{1}$ and \textit{K}$_{2}$ obtained by VSM data were matched to MPMS data at $300$~K.}
\label{7Tki}
\end{figure}
The magnetocrystalline anisotropy field was determined to be $\sim 10$~T (T = 2 K) and $\sim 6$~T (T = 300 K) for a Ce$_{1.66}$Mg$_{1.34}$Co$_{9}$ single crystalline sample, as shown in FIG.~\ref{Anisotropyfieldfu}. The anisotropy field was determined by the linear extrapolation of the observed moment along the plane up to the saturation moment.   

\begin{figure}
\includegraphics[scale =0.35]{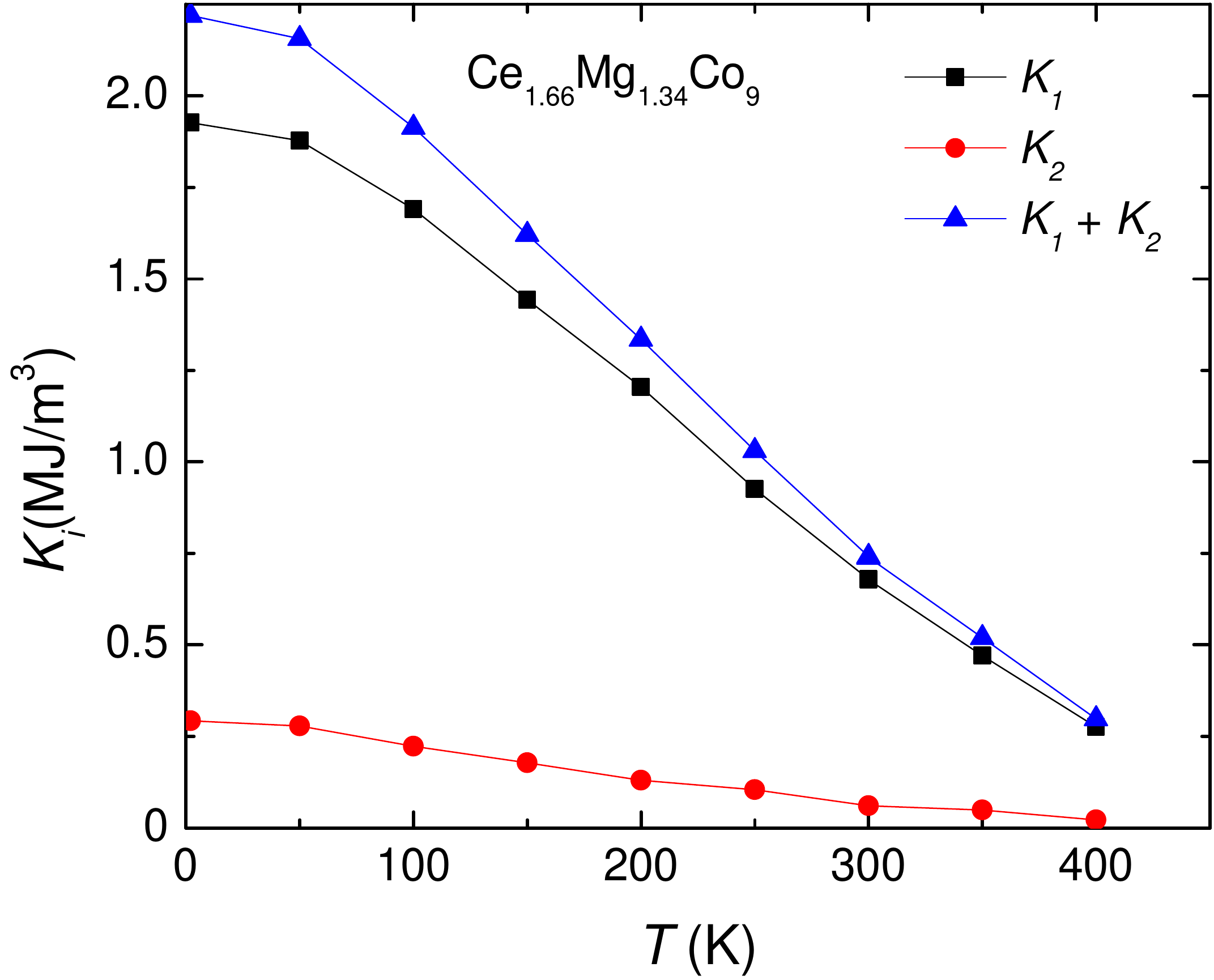}
\caption{Temperature dependent anisotropy energy constants for Ce$_{1.66}$Mg$_{1.34}$Co$_{9}$. The anisotropy constants were determined from the Sucksmith-Thompson plot. The K$_1$ and K$_2$ values up to $300$~K were measured using MPMS and higher temperature data are measured using VSM. }
\label{K1K2scaled}
\end{figure} 
 
The anisotropy energy was quantified by using the Sucksmith-Thompson plot for  field dependent magnetization data along the plane as shown in FIG. ~\ref{7Tki}. In the Sucksmith-Thompson plot, the ratio of magnetizing field with hard axis magnetization data is related to the anisotropy constants $K_{1}$ and $K_{2}$, saturation magnetization $M_{s}$ and hard axis magnetization ($M_\perp$) as shown in equation (1)\citep{Sucksmith362, VTaufourMnBi2015}. 
 
\begin{eqnarray}
\frac{\mu_0 H}{M_\perp} = \frac{2K_{1}}{M_{s}^2} + \frac{4K_{2}}{M_{s}^4}M_\perp^2
\end{eqnarray}

The intercept of the Sucksmith-Thompson plot gives the anisotropy constant $K_{1}$ and slope gives the anisotropy constant $K_{2}$. FIG.~\ref{K1K2scaled} shows the temperature variation of the measured anisotropy constants. There was a slight mismatch ($<2\%$) in the anisotropies data above and below the $300$~K obtained from VSM and MPMS data. The VSM data were scaled to MPMS data at $300$~K since slope of Sucksmith-Thompson plot are better determined with higher applied field. Here the total anisotropy energy of Ce$_{1.66}$Mg$_{1.34}$Co$_{9}$ is determined to be $2.2$~MJ/m$^3$. Such high anisotropy energy density makes the Ce$_{1.66}$Mg$_{1.34}$Co$_{9}$ as a potential candidate for permanent magnet applications.

\section{Conclusions}
We investigated the effect of Mg substitution into the  Ce$_3$Co$_{9}$ (e.g. CeCo$_3$) binary phase where Mg partially replaces the Ce atom on the 6\textit{c} crystallographic site giving rise to the Ce$_{3-x}$Mg$_{x}$Co$_{9}$ solid solution for $0\leq x \lesssim 1.4$. The substituted Mg induces ferromagnetism; the Curie temperature of the solid solution increases with higher content of Mg and becomes maximum ($450$~K) for Mg content \textit{x} = 1.35. The magnetic anisotropy was determined for a self-flux grown Ce$_{1.66}$Mg$_{1.34}$Co$_{9}$ single crystal. The uniaxial-anisotropy field was determined to be~$\sim10$~T at $2$~K and $\sim6$~T at $300$~K. The anisotropy energy density was determined to be $2.2$~MJ/m$^3$ at $2$~K.  With these observed magnetic properties, Ce$_{3-x}$Mg$_{x}$Co$_{9}$ solution shows a potential to be used as a permanent magnet.

\section{Acknowledgements}
\noindent
We would like to thank Dr. T. Kong for useful discussions. Dr.  Warren Straszheim is acknowledged for doing SEM on various samples. This research was supported by the Critical Materials Institute, an Energy Innovation Hub funded by the U.S. Department of Energy, Office of Energy Efficiency and Renewable Energy, Advanced Manufacturing Office. Q.L. is supported by the office of Basic Energy Sciences, Materials Sciences Division, U.S. DOE. This work was performed at the Ames Laboratory, operated for DOE by Iowa State University under Contract No. DE-AC02-07CH11358.\\*

\bibliography{TejResearch-R2Co9Mg}
\end{document}